\documentclass[twocolumn]{aastex631}

\newcommand\msun{$M$\mbox{$_{\normalsize\odot}$}}

\newcommand\lbol{$L$\mbox{$_{\rm bol}$}}

\newcommand\logl{$\log (L/L_\odot)$}

\newcommand\bchsti{{\it BC$_{F814W}$}}

\usepackage{soul}
\newcommand\N[1]{{\color{teal} \bf #1}} 
\newcommand\J[1]{{\color{violet} \bf #1}} 

\begin{document}

\title{The red supergiant progenitor luminosity problem}

\author[0000-0003-4666-4606]{Emma R.\ Beasor}\altaffiliation{Bok Fellow}
\affil{Steward Observatory, University of Arizona, 933 North Cherry Avenue, Tucson, AZ 85721-0065, USA}

\author[0000-0001-5510-2424]{Nathan Smith}
\affil{Steward Observatory, University of Arizona, 933 North Cherry Avenue, Tucson, AZ 85721-0065, USA}

\author[0000-0001-5754-4007]{Jacob E.\ Jencson}
\affil{IPAC, California Institute of Technology, 1200 East California Boulevard, Pasadena, CA 91125, USA}

\defcitealias{mcdonald2022red}{M22}


\begin{abstract}
Analysis of pre-explosion imaging has confirmed red supergiants (RSGs) as the progenitors to Type II-P supernovae (SNe). However, extracting the RSG's luminosity requires assumptions regarding the star's temperature or spectral type and the corresponding bolometric correction, circumstellar extinction, and possible variability. The robustness of these assumptions is difficult to test, since we cannot go back in time and obtain additional pre-explosion imaging. Here, we perform a simple test using the RSGs in M31, which have been well observed from optical to mid-IR. We ask the following: By treating each star as if we only had single-band photometry and making assumptions typically used in SN progenitor studies, what bolometric luminosity would we infer for each star? How close is this to the bolometric luminosity for that same star inferred from the full optical-to-IR spectral energy distribution (SED)?  We find common assumptions adopted in progenitor studies systematically underestimate the bolometric luminosity by a factor of 2, typically leading to inferred progenitor masses that are systematically too low. Additionally, we find a much larger spread in luminosity derived from single-filter photometry compared to SED-derived luminosities, indicating uncertainties in progenitor luminosities are also underestimated.
When these corrections and larger uncertainties are included in the analysis, even the most luminous known RSGs are not ruled out at the 3$\sigma$ level, indicating there is currently no statistically significant evidence that the most luminous RSGs are missing from the observed sample of II-P progenitors. The proposed correction also alleviates the problem of having progenitors with masses below the expected lower-mass bound for core-collapse. 
\end{abstract}


\section{Introduction} \label{sec:intro}
The most robust test of stellar evolutionary theory for massive stars comes from linking observed supernovae (SNe) directly to their detected progenitors. It has long been suggested that all single stars with initial masses between 8--30\msun\ \citep[e.g.][]{meynet2000stellar} will pass through the red supergiant (RSG) phase, before ending their lives as Type II-P core-collapse supernovae (CCSNe). It was also expected that above some initial mass ($\sim$30 \msun) the cores of stars are so massive that infall after core collapse will quench a successful SN and will lead to the creation of a black hole. However, both the analysis of SN progenitor detections or upper limits \citep[e.g.][]{smartt2009death} and some evolutionary models \citep[e.g.][]{sukhbold2018high} now suggest the upper mass limit for II-Ps may in fact be lower ($\sim$20\msun). 

Following after the first case of SN~1987A, serendipitous pre-explosion observations of several SN progenitors \citep[see][and references therein]{smartt2015observational} provided important and direct information about the endpoints of stellar evolution for massive stars.  Namely, it was through the direct imaging method that RSGs were confirmed as the long-expected progenitors of Type II-P SNe \citep[e.g.][]{maund2004massive}. However, this evidence also pointed to an apparent discrepancy between the presumed initial mass distribution of RSGs in the field compared to RSGs dying as supernovae. \citet{smartt2009death} argued that the apparent discrepancy between the initial mass distribution of RSGs in nearby field populations, as compared to the initial mass distribution of Type II-P progenitors, was clear evidence that some RSGs do not die as bright SNe II-P. In particular, RSGs with inferred initial masses of 20-30\msun \, are clearly seen in nearby stellar populations, but the maximum observed SN~II-P progenitor mass is only $\sim$ 17\msun. 

The unknown fates of stars in the 20-30\msun\ range has famously been termed the ``Red Supergiant Problem'', and has led to an explosion of theories that may explain the phenomenon. One leading idea is that these stars fail to produce bright SN explosions as their cores collapse into a black hole. Theoretical studies also seem to be converging on an upper mass limit for II-Ps of around 20\msun, see e.g. \citep{sukhbold2018high}. However, the inferred distribution of RSG progenitor initial mass has some unresolved issues.  For one, even if the distance to the object is known, there is considerable uncertainty stemming from the assumption of spectral type (SpT) for each progenitor star. \citet{davies2018initial} showed that assuming a constant SpT of M0 for the RSG progenitors of SN~II-Ps could lead to an underestimation of the initial mass of 5\msun. By taking this into account, the significance of the missing RSGs was reduced to $<$2$\sigma$. In addition, converting luminosities to initial masses is heavily dependent on the mass-luminosity relation chosen \citep[see Table 2 within ][]{davies2018initial} furthering the uncertainty. One way to circumvent these uncertainties is to skip a step altogether. Rather than comparing the inferred mass distribution of the field RSGs to inferred masses of the progenitor RSGs, one can simply compare the measured {\it luminosity} distributions directly. In this way, considerable uncertainties inherent to stellar evolutionary models can be removed from the problem.

When comparing field RSGs to SN progenitors in terms of luminosity, it appears there {\it is} a dearth of Type II-P progenitors above \logl=5.2, in tension with the empirical luminosity limit for RSGs of \logl=5.5 \citep{davies2018humphreys}. In \citet{davies2020red} it was shown, however, that the absence of high luminosity progenitors can in part be explained by the steepness of the $L$-distribution and low number statistics, and they estimated that the significance of the ``missing'' progenitors is between 1$\sigma$ and 2$\sigma$. Further complications arise when taking into account the effect of circumstellar dust which may have been destroyed by the SN \citep[e.g.][]{walmswell2012circumstellar}.

However, even when eliminating systematic errors from the conversion of luminosity to initial mass, there is still a glaring uncertainty that mars the pre-explosion imaging technique; it is impossible to know the true shape of the progenitor spectral energy distribution (SED) from only a single-band detection. This means we are forced to make assumptions about the intertwined values for reddening (foreground and circumstellar) and temperature. Here, we conduct an experiment to characterize the potential systematic effects, by directly comparing the inferred luminosities from single HST $I$-band (F814W) photometry -- the filter in which serendipitous pre-explosion imaging is most frequently found -- to the bolometric luminosity found when integrating under the entire SED (a method which is free from model assumptions) {\it for the same sample of stars}. To do this we use a sample of RSGs from M31 \citep{mcdonald2022red} that have well-sampled SEDs from optical to the mid-IR, and in addition, have HST $F814W$-band observations from the Panochromatic Hubble Andromeda Treasury (PHAT) survey \citep{dalcanton2012ophat}. With this direct comparison, we will demonstrate the potential systematic effects on the luminosities determined for SNe when only a single-band observation is available. 

In Section \ref{section:method} we detail the methods used to determine the luminosities of each RSG in the sample. In Section \ref{section:results} we compare the luminosties determined for each RSG via the BC and SED methods and discuss potential implications for SN progenitors. We also apply the results of this work to the Red Supergiant Problem. In Section 4 we summarise the results and make suggestions for future work on RSG progenitors. 

\defcitealias{mcdonald2022red}{M22}
\section{Method}\label{section:method}
\begin{figure}
    \centering
    \includegraphics[width=\columnwidth]{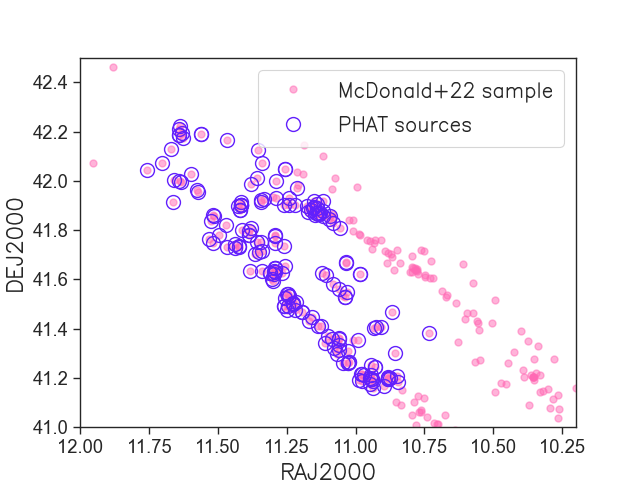}
    \includegraphics[width=\columnwidth]{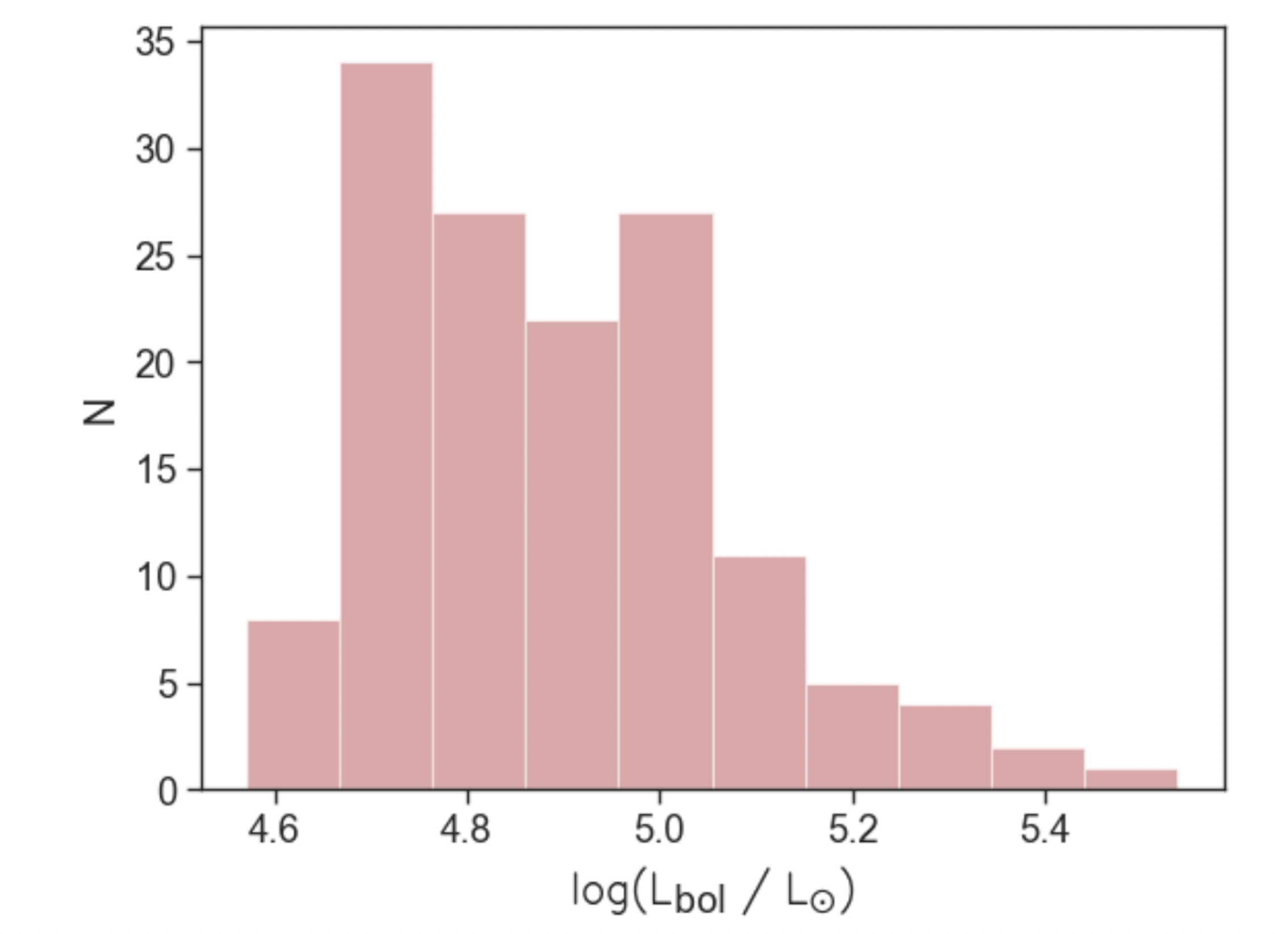}
    \caption{{\it Top panel: }Sources in the \citetalias{mcdonald2022red} catalogue (pink filled circles) cross matched with sources in PHAT (purple open circles). {\it Bottom panel: }Luminosity distribution for the cross matched sample. }
    \label{fig:matchup}
\end{figure}

We begin by matching the stars in the \citet[][hereafter M22]{mcdonald2022red} catalog to point sources in the PHAT survey using {\tt astropy} routine {\tt SkyCoord}. In Fig. \ref{fig:matchup} we show the positions of the stars, where the pink circles show sources from \citetalias{mcdonald2022red} and those which correspond to a PHAT source highlighted with a blue circle. In some cases we found multiple PHAT sources associated with RSGs from the \citetalias{mcdonald2022red} sample, implying possible binary/multiple systems or close neighbors as projected on the sky. For simplicity, we ignore any objects with multiple matching PHAT sources. After excluding these multiples, we found 139 sources within PHAT that are likely associated with RSGs in the \citetalias{mcdonald2022red} sample (see the bottom panel of Fig. \ref{fig:matchup}). 

We now compare the luminosities of each of the RSGs derived via two methods. The SED derived luminosities were calculated in \citetalias{mcdonald2022red} by using {\tt IDL} routine {\tt int$\_$tabulated} and adopting an M31 distance modulus of 24.4 \citep{karachentsev2004m31}. In this method, the luminosity is obtained by simply integrating under the observed SED. The only assumption necessary for this method is that any of the flux lost at shorter wavelengths due to absorption by circumstellar material (CSM) is re-emitted at longer wavelengths, and so when integrating from optical to mid-IR wavelengths, all of the stars' flux is recovered (see DCB18).

The second method we use to obtain a luminosity is that which is commonly used for pre-explosion imaging of SN progenitors. Pre-explosion imaging relies on serendipitous detections or upper limits from surveys that often used only a single band, most commonly the HST $F814W$ ($I$-band) filter \citep[see e.g. Table 4 within][]{davies2018initial}. A number of assumptions are therefore needed to derive a luminosity. We begin by converting the apparent I-band magnitude to an absolute magnitude by applying a distance modulus of 24.4 (see above) and de-reddening the photometry according to the extinction value from PHAT, see \citepalias{mcdonald2022red}. We then apply the same bolometric correction as adopted in the study by \citet{smartt2009death} (\bchsti = 0.9) to convert the I-band magnitude to a bolometric magnitude. Finally, we convert this to luminosity. 

\begin{figure}
    \centering
    \includegraphics[width=\columnwidth]{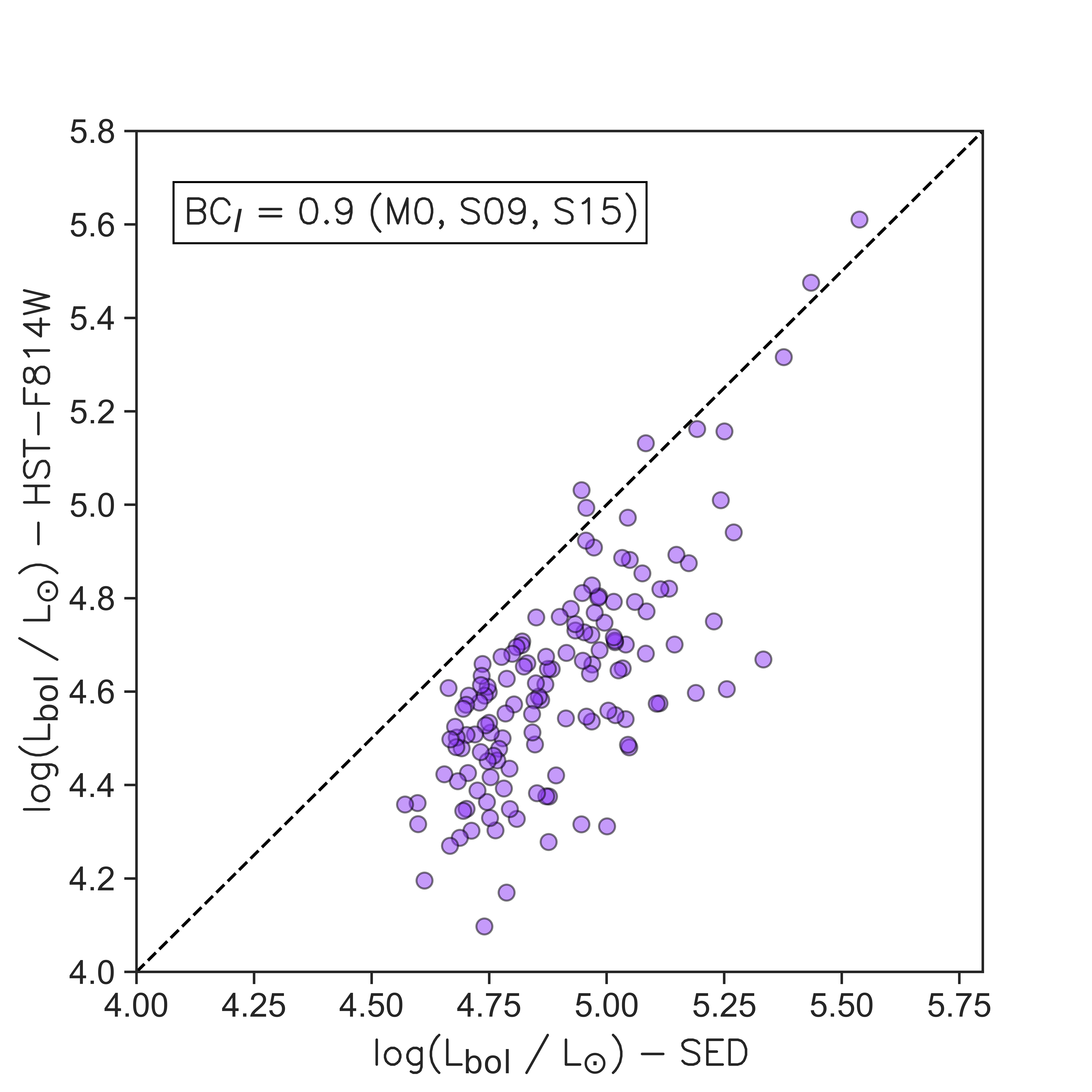}
    \includegraphics[width=\columnwidth]{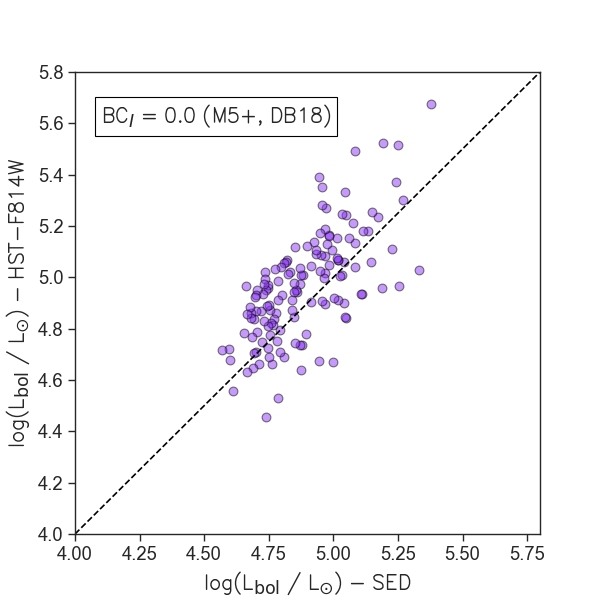}
    \caption{Comparison between luminosities derived by integrating under the observed SED and the luminosities derived from the HST $F814W$ photometry. {\it Top panel:} Using \bchsti = 0 from \citet{smartt2009death}. {\it Bottom panel:} Using \bchsti = 0.9 from \citet{davies2018initial}.}
    \label{fig:comp}
\end{figure}

\begin{figure}
    \centering
    \includegraphics[width=\columnwidth]{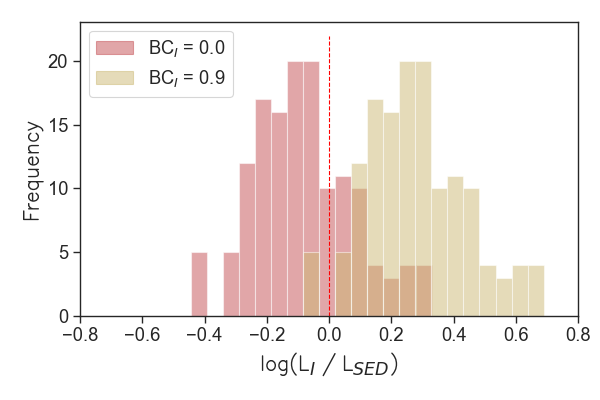}
    \caption{Histograms showing the values for the log ratio of the luminosity determined from the I-band method using the bolometric corrections for early (BC = 0.9) and late (BC = 0) type RSGs. }
    \label{fig:ratio}
\end{figure}

The comparison between the SED-derived luminosities and the $I$-band derived luminosities (using \bchsti = 0.9 mag; \citealt{smartt2009death}) is shown in the top panel of Fig. \ref{fig:comp}.  \citet{davies2018initial} argued that RSGs tend to evolve to later spectral types of M5 before their death. We therefore also show the results when assuming a later spectral type (and thus smaller \bchsti) as suggested by \citet{davies2018initial}, in the bottom panel of Fig. \ref{fig:comp}, using \bchsti = 0.0 mag. We demonstrate the differences again in Fig \ref{fig:ratio}, where we show the ratio L(I-band)/L(SED) for each BC value.

\section{Results and Discussion}\label{section:results}
Figure 2 demonstrates that when treating the RSGs in M31 the same way that SN progenitors are often treated, i.e. assuming a SpT and hence BC, there can be significant disagreement between this derived luminosity and the luminosity as derived more accurately by the SED method. 

When adopting the same assumptions applied to pre-explosion photometry in the original `RSG Problem' papers \citep[][\bchsti = 0.9]{smartt2009death,smartt2015observational}, we find that luminosities from single-band photometry systematically {\it underestimate} the bolometric luminosity of the star by about a factor of 2 (i.e. by an average of 0.3 dex in Fig 2). This translates  to a shift in the derived initial mass of $\sim$ 4 \msun, see e.g. Fig 4 within \citep{smartt2009death}. Importantly, this indicates that the distribution of initial masses for SN II-P progenitors summarized by \citet{smartt2009death} should be shifted to higher mass, which will modify both the lower bound and upper bound for the inferred range of masses of SN II-P progenitors.   On the other hand, when applying a BC appropriate for stars of a later spectral type (\bchsti = 0), the average offset between the two methods is much smaller (-0.08 dex), although there is still dispersion on the order of 0.2 dex.  We now discuss the potential reasons for this discrepancy between SED-derived luminosities and $F814W$-derived luminosities, as well as the effect this may have on progenitor detections.

\subsection{Potential causes of discrepancy}
There are likely two main causes for the discrepancy between the SED method and the single-band method: the uncertainty of the SpTs for each star and stellar variability.\footnote{ We note that most RSGs do not appear to be significantly extincted by their circumstellar dust, see e.g. \citet{beasor2022dusty}.} In this case, we know that most of the RSGs in the M31 sample are not close to the time of explosion and therefore are expected to span a broader range of SpTs than SNe progenitors; this is because SN progenitors should all be located the coolest and most luminous endpoints of their evolutionary tracks, whereas the population of RSGs in M31 sample a range of points along those tracks. For this  reason, the assumed value of BC=0.0 mag causes a slight overestimate of the luminosity compared to the SED value (Fig 3).  While we do not know their individual SpTs, we do know that the range of SpTs is broad \citep[M0-M5, see e.g.][]{humphreys1988m31,massey2016m31}, while SN~II-P progenitors likely all have later SpTs (DB18). If we allow \bchsti\ to vary, while still using a single BC for the entire RSG population in M31, we find that a BC of 0.2 mag \citep[corresponding to an average SpT = M4, see][]{davies2018initial}  yields the smallest offset between the two methods, implying the sample of RSGs in M31 have preferentially late spectral types despite not all being close to SN. We suggest the relation between SpT and BC may need further refinement, this will be addressed in a future work. While this reduces the average offset, there is still a dispersion at a level of 0.2 dex around the 1:1 relation.

\begin{figure*}
    \centering
    \includegraphics[width=\columnwidth]{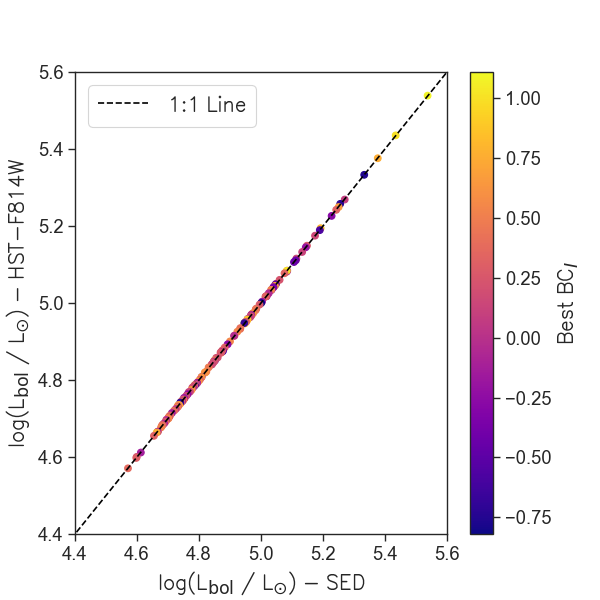}\includegraphics[width=\columnwidth]{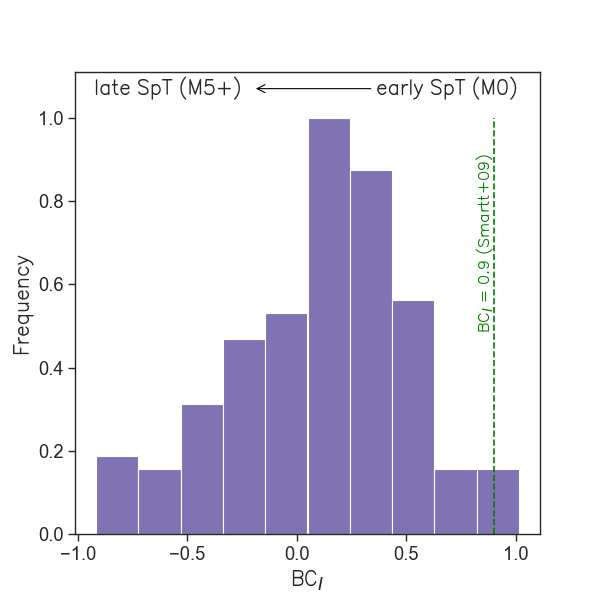}
        \caption{Required BC$_{\rm I}$ values to make the luminosities from the $I$-band method match the SED method. }
    \label{fig:varybc}
\end{figure*}

We now investigate the range of BC values that would be required to achieve a perfect match between the luminosities determined via the SED and the single-band methods. When leaving \bchsti\ as a free parameter for each individual star, it is possible to find agreement between the two luminosity methods using BC values ranging between -1 and 1 (see Fig. \ref{fig:varybc}). BC values less than 0 would imply a very late spectral type (DB18) or a local reddening correction larger than the adopted value. 

\begin{figure*}
    \centering
    \includegraphics[width=\columnwidth]{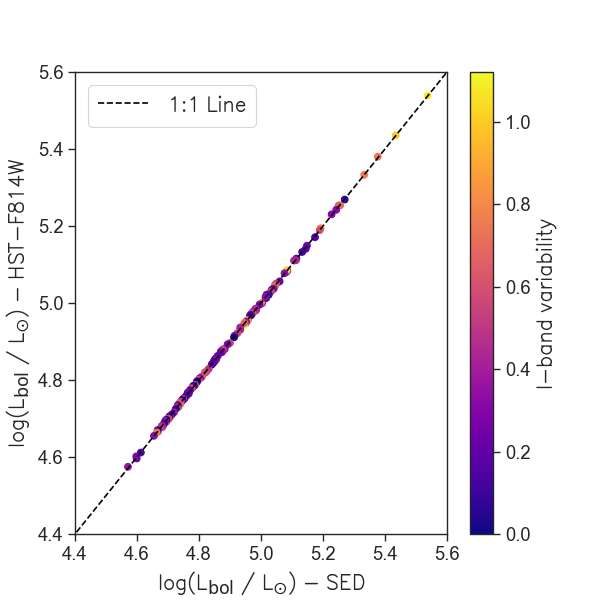}\includegraphics[width=\columnwidth]{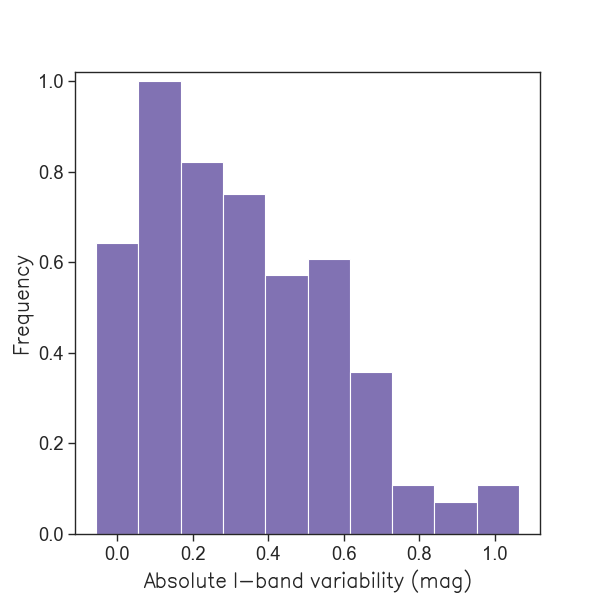}
    \caption{Level of I-band variability required to minimise the discrepancy between the BC and SED luminosity calculation methods. }
    \label{fig:varplot}
\end{figure*}

The other potential cause for the large spread is stellar variability. It is known that RSGs are variable \citep[e.g.][]{soraisam2018} by up to 1 mag in the optical, decreasing to around 0.1 mag at mid-IR wavelengths. For luminosity estimates that rely on single-band photometry, this could potentially have a large effect on the derived \lbol. Since multi-epoch SN progenitor detections are still rare, it is usually not possible to tell how variable a star was, or at what point in its variability cycle the star was at when the observations were taken. Some studies do search for variability in SN progenitors \citep[][]{neustadt2023, jencson2023,jacobson2022final}. For the stars in M31, the photometry used in the SED method is all ground-based, and not taken contemporaneously with the HST PHAT data. The inherent variability within the RSGs could potentially influence the dispersion in luminosity observed within the sample, but the overall average should remain unaffected, as it is unlikely that the detections have been biased toward any particular phase of the variability cycle. 
In the same way as above, we now investigate the level of variability that would be required to allow the luminosities from each method to match perfectly, see Fig. \ref{fig:varplot}. We find that by adding variability of up to $\pm$ 1 mag to the F814W photometry it is possible to reconcile the inconsistencies between the SED and the I-band luminosity method.

As such, we suggest that the offset is likely caused by a combination of uncertain BCs and stellar variability. It is also important to note that the RSGs in M31 represent a best-case scenario compared to most SN progenitor detections, since the distance and foreground extinction are well known. More specifically, the line-of-sight interstellar extinction towards progenitors is determined from a combination of Milky Way reddening derived from dust maps and possible host galaxy extinction determined from e.g Na I D lines seen in the SN spectrum. However, host galaxy reddening determined from observations of SN light does not account for possible CSM dust around the progenitor that may have since been destroyed in the SN explosion itself. The uncertainty introduced by this is inherently in one direction - i.e. it potentially makes the progenitor more luminous if there was CSM dust. 

\subsection{Impact on SN progenitor luminosities}
Linking progenitor stars to their explosive deaths is a clear and direct test of stellar evolutionary theory. However, the results here highlight the difficulty in deriving fundamental progenitor properties from single-band imaging. Namely, both the precision and accuracy in deriving a progenitor luminosity and initial mass from optical photometry are significantly worse than previously assumed.

The situation is, however, even worse than demonstrated above. In this work, we are working under idealised conditions where the foreground extinction and distance to M31 are well known \citep{dalcanton2012ophat}, and despite this, there is still significant uncertainty in the luminosities derived from single-bands. In real situations with SN progenitors, the uncertainties on distance and extinction are likely far higher, in addition to the uncertainty surrounding the SpT and any intrinsic variability. 

Indeed, it is difficult to determine SpTs (and hence the appropriate BCs) for stars without spectra. \citet{smartt2009death} assumed all RSG progenitors would be the earliest spectral type, M0, and hence would require a \bchsti = 0.9 \citep[taken from ][]{elias1985m}. Here, we have shown that when treating the M31 RSGs the same way, this assumption artificially causes an underestimation of a star's luminosity by an average of 0.3 dex. 

Real RSGs have a range of later M types, especially in the case of SN progenitors, and this discrepancy is improved somewhat when using a BC for a later SpT. \citet{davies2018initial} showed that RSGs likely evolve to later spectral types as they approach SN, and suggest that instead for SN progenitors we should assume a later SpT (in absence of any further knowledge). When reanalysing Type IIP progenitor luminosities, a smaller BC appropriate for late type (M5+) RSGs \citet{davies2018initial} showed that on average the inferred luminosity (and hence initial mass) increased slightly. Here, for the RSGs in M31, we have shown that while the average offset between the SED and the I-band method is improved with a more appropriate BC, there is still scatter on the order of 0.2 dex (see bottom panel of Fig. \ref{fig:comp}. While the population of stars in M31 is likely not directly comparable to SNe progenitors, since not all of the RSGs will be close to their time of death, the results here suggest the errors on luminosities determined using single-band photometry are likely larger than previously assumed. 

Variability in SN progenitors may also have a significant impact on the luminosity determined. Here, we have shown by assuming I-band variability up to $\pm$ 1 magnitude, the dispersion of the luminosities can be reduced. 

Combined, intrinsic variability of the RSGs and the uncertainty in bolometric correction for the progenitor sources likely result in an error on luminosity that is larger than currently assumed. This has an impact at the high and the low end of the progenitor luminosity function, and may go some way to explain the lack of high-mass SN II progenitors as well as the presence of stars with luminosities below \logl\ = 4.5 \citep{fraser2014ej}. The lower luminosity progenitors may form a more severe conflict with theory than the missing high-mass progenitors. For a single star to undergo core-collapse, its initial mass must be above $\sim$ 8\msun\ \citep[e.g.][]{heger2003how}. Below this initial mass, single stars are expected to die as a planetary nebula and white dwarf. Binary evolution processes such as mass transfers or mergers can give a tail of core-collapse SNe at later ages than a single 8.5\msun\ star (and hence, we may see ccSN explosions at {\it  initial} masses lower than M = 8.5\msun, \citealp{zapartas2017, zapartas2019, zapartas2021effect}), but the progenitor stars themselves should have a luminosity similar to a single star of $M_{\rm ini}$ $>$8 \msun\ star at the time of explosion, since they have gained mass throughout their evolution. There have been progenitor detections that, when using standard assumptions to estimate luminosity, appear to have luminosities below this limit (e.g. SN 2003gd, SN 2005cs, see Fig. {\ref{fig:clds}}). The results found here alleviate this tension since they imply the luminosities for these low-mass progenitors likely have significantly larger errors than previously assumed. In addition, both of these progenitors have their luminosities estimated from F814W photometry only, and as discussed in Section \ref{sec:midIR} by progenitors with F814W and mid-IR photometry, relying on F814W photometry only may lead to the luminosity being significantly underestimated.   

\subsubsection {Progenitors with mid-IR pre-explosion photometry}\label{sec:midIR}
Only a handful of SN progenitors have pre-explosion detections in the mid-IR beyond 2~$\mu$m. 
An interesting case is SN 2017eaw. The progenitor was identified as being a dusty RSG from multiwavelength pre-explosion photometry \citep{kilpatrick2018,vandyk2019}, including near and mid-IR data from Spitzer out to 4.5\,$\mu$m and two epochs of HST/F814W. \citet{kilpatrick2018} derived a luminosity from SED fitting of \logl\ $= 4.9$, assuming a distance to NGC\,6946 of 6.7~Mpc \citep[$\mu = 29.13$][]{tikhonov2014} and total foreground extinction of $E(B-V) = 0.34$~mag. \citet{vandyk2019} find a larger luminosity of \logl\ $= 5.1$, but assumed a correspondingly larger distance of 7.73~Mpc from their own TRGB analysis. 
We now take only the HST/F814W photometry and compare the luminosity that would have been found for the progenitor had there been no other photometry available.

For SN 2017eaw, there were two available epochs of $F814W$ data from 2004 and 2016. We begin by taking the 2004 photometry from \citep{vandyk2019}, $m_{\rm F814W}$ = 22.60 mag, and determine the luminosity adopting the same values for the distance and extinction, and find \logl\  = 4.8 and 4.5 using \bchsti\ = 0 mag (M5) and \bchsti\ = 0.9 mag (M0), respectively. When using the photometry from 2016 of $m_{\rm F814W}$ = 22.8 mag, the corresponding luminosity estimates decrease to \logl\ = 4.7 and 4.4. This results in offsets from the SED-derived luminosities of 0.3--0.4~dex for an M5 BC and significantly larger offsets of 0.7-0.8~dex for an M0 BC. The F814W photometry reported by \citet{kilpatrick2018} are $\sim$0.05~mag brighter at both epochs. Conducting the same exercise, we find offsets from their SED-derived luminosities of 0.1--0.2~dex for an M5 BC and  0.5-0.6~dex for an M0 BC.

Recently, SN 2023ixf was detected in the nearby spiral M101 \citep{itagaki2023}, a galaxy that has been imaged many times across multiple wavelengths. Several works have examined the extensive pre-explosion dataset---including optical imaging from HST and ground-based resources, ground-based near-IR imaging, and Spitzer mid-IR imaging---and derived SED-based luminosities of the RSG progenitor spanning a wide range: \logl\ $= 4.74 \pm 0.07$ at the low end \citep{kilpatrick2023}, mid-range values of \logl\ $= 4.8$--5.0 \citep{neustadt2023,vandyk2024,xiang2024a}, and high-end estimates of \logl\ $\approx 5.1$ \citep{jencson2023,niu2023,ransome2024,qin2023}. Notably, \citet{pledger2023} examined only the HST data, inferring an $I$-band absolute magnitude of  $M_{F814W} = -5.1^{+0.7}_{-0.5}$, which corresponds to \logl\ $=3.6$ (M0) or \logl\ $=3.9$ (M5). On the other end, \citet{soraisam2023} obtained \logl\ $= 5.37 \pm 0.12$ by applying the RSG period--luminosity relation of \citet{soraisam2018} to the observed pre-explosion periodicity. Despite the large range in luminosity estimates, there has been general agreement that the progenitor was highly dust-enshrouded and undergoing enhanced pre-explosion mass-loss, essentially guaranteeing that luminosities derived solely from optical data will be vast underestimates. 

We now repeat the exercise of taking only the HST/F814W photometry to derive a luminosity estimate for SN\,2023ixf. We adopt $m_{F814W} = 24.31 \pm 0.05$ mag (taken on 2002 Novemeber 16) and a total foreground extinction of $E(B-V) = 0.04$ mag from \citep{vandyk2024}, and a distance of $6.85 \pm 0.32$~Mpc ($\mu = 29.18$; \citealp{riess2022}). Thus, we find \logl\ $\approx3.5$ (M0 BC) or \logl\ $=3.9$ (M5 BC), consistent with the comparable analysis of \citet{pledger2023}. As expected, the offsets from the SED-based luminosities are remarkably large, 1.2--1.6~dex for an M0 BC and 0.8--1.2~dex for an M5 BC.  

Finally, SN\,2024ggi was recently discovered in the nearby galaxy NGC\,3621 \citep{tonry2024}, again with extensive pre-explosion coverage in the optical, near- and mid-IR. So far, \citet{xiang2024b} has analyzed this dataset and derived an SED-based luminosity of \logl\ $= 4.9^{+0.05}_{-0.04}$. They also determined $M_{F814W} = -6.21 \pm 0.08$ mag (assuming total foreground extinction $E(B-V) = 0.19$~mag, \citealp{zhang2024}, and distance modulus $\mu = 29.14 \pm 0.06$~mag, \citealp{tully2013}). Based on the $I$-band magnitude alone, we would thus infer \logl\ $=4.0$ (M0) and $4.4$ (M5), once again giving substantial offsets from the SED-based estimates of 0.9 and 0.5~dex, respectively.  

Overall, these three SN progenitor cases --- where we have sufficient multiwavelength photometry to place the best available constraints on the SED and hence the true luminosity --- illustrate that the method of applying a reasonable assumed bolometric correction to single-band photometry generally does a poor job of approximating the true luminosity.  In particular, underestimating the luminosity by a factor of 10 or more is not unusual using this method\footnote{We note that the three SN prognitors discussed in this work may not be representative of the entire RSG progenitor sample, as 2023ixf and 2024ggi are early `flash' SNe, implying they are particularly dusty objects. It is not clear how many progenitors had mid-IR coverage (e.g. Spitzer) but remained undetected.}. This should raise concern about luminosity and initial mass estimates for other SN progenitors where such multiwavelength information is not available.

\subsection{The Red Supergiant Problem}
Since \citet{smartt2009death} postulated that the highest mass (or luminosity) RSGs may not end their lives as SNe, various evolutionary scenarios have been proposed to account for the RSG problem, such as enhanced mass-loss \citep{georgy2012yellow}, direct collapse to black hole without a bright SN \citep{sukhbold2018high}, or massive RSGs producing SN subtypes such as SNe IIn or II-L instead of SNe II-P \citep{smith2011observed,smith09vy}. It has also been suggested that the `missing' RSGs may be an artifact of unaccounted extinction \citep{walmswell2012circumstellar,smith2011observed} or incorrect bolometric corrections \citep{davies2018initial}. Whatever the explanation, the statistical significance of the result has, however, remained contentious in the literature \citep[see e.g.][]{davies2020red, davies2020on, kochanek2020on}. 

Here we have demonstrated that when RSG luminosities are determined by single F814W-band photometry, they are both (1) systematically underestimated by a factor of 2 when adopting standard BC values, and (2)  have larger errors than previously claimed, due to the added uncertainty surrounding a progenitors' variability and SpT. The work presented here indicates that the potential additional errors on luminosities determined from HST I-band detections could be as high as 0.2 dex. 

\begin{figure*}
    \centering
    \includegraphics[width=15cm]{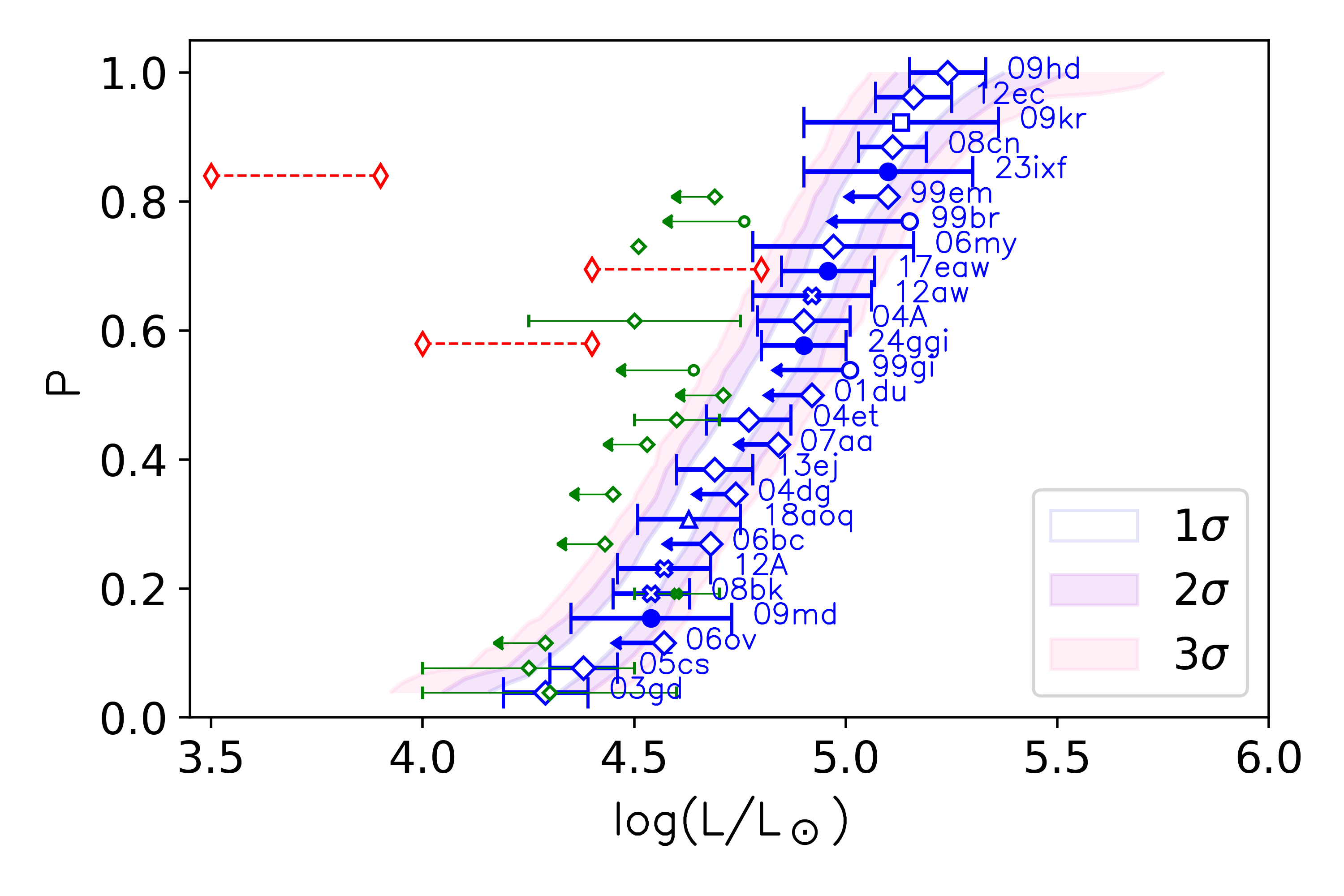}
    \includegraphics[width=15cm]{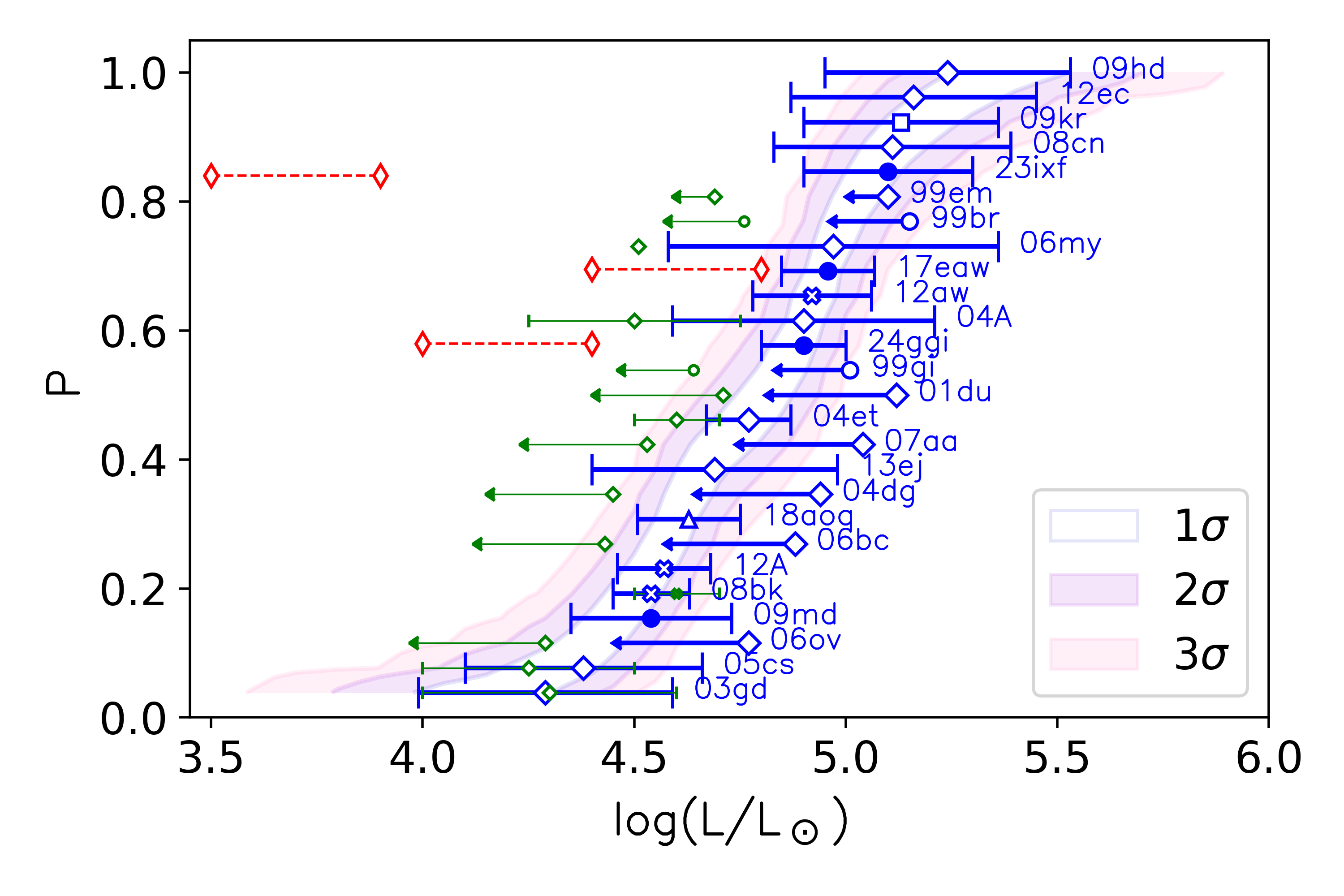}
    \caption{Cumulative luminosity distributions for Type II progenitors with standard errors (top panel) and additional systematic uncertainty of 0.2 dex added to I-band observations (bottom panel). In each case, the symbols represent the following; filled circles - multi-waveband, open squares - F555W, open diamonds - F814W, open crosses - K-band. We also show the luminosity measurements from \citet{smartt2009death} in via the green symbols. One might expect that the older estimates from Smartt et al. (green) should all be fainter by 0.36dex as compared to the revised estimates (blue).  While the newer estimates are corrected by this factor, they also have updated photometry, distance, and extinction estimates as detailed in DB18, so they are not all offset by the same amount. Finally, we show the luminosities that would have been estimated for SN 2017eaw, SN 2023ixf, and SN 2024ggi if only I-band photometry were available using the BCs for both an M0 and an M5 RSG via the red diamonds.}
    \label{fig:clds}
\end{figure*}

We now explore the effect this additional uncertainty may have on the statistical significance of the RSG problem. To do this, we repeat the analysis presented in \citet{davies2020red}, including SN progenitors that have been discovered more recently, and include an additional error on any SN progenitor detected in the F814W filter. To include the additional error on each of these \lbol\ measurements we randomly sample from a Gaussian centered on zero with a standard deviation of 0.2 dex, and we add this to the normal errors. We repeat this 10$^{4}$ times and the luminosity that is used in analysis is taken to be the average with errors drawn from the standard deviation. The results of each run are shown in Figure \ref{fig:clds}.

\begin{figure*}
    \centering
    \includegraphics[width=\columnwidth]{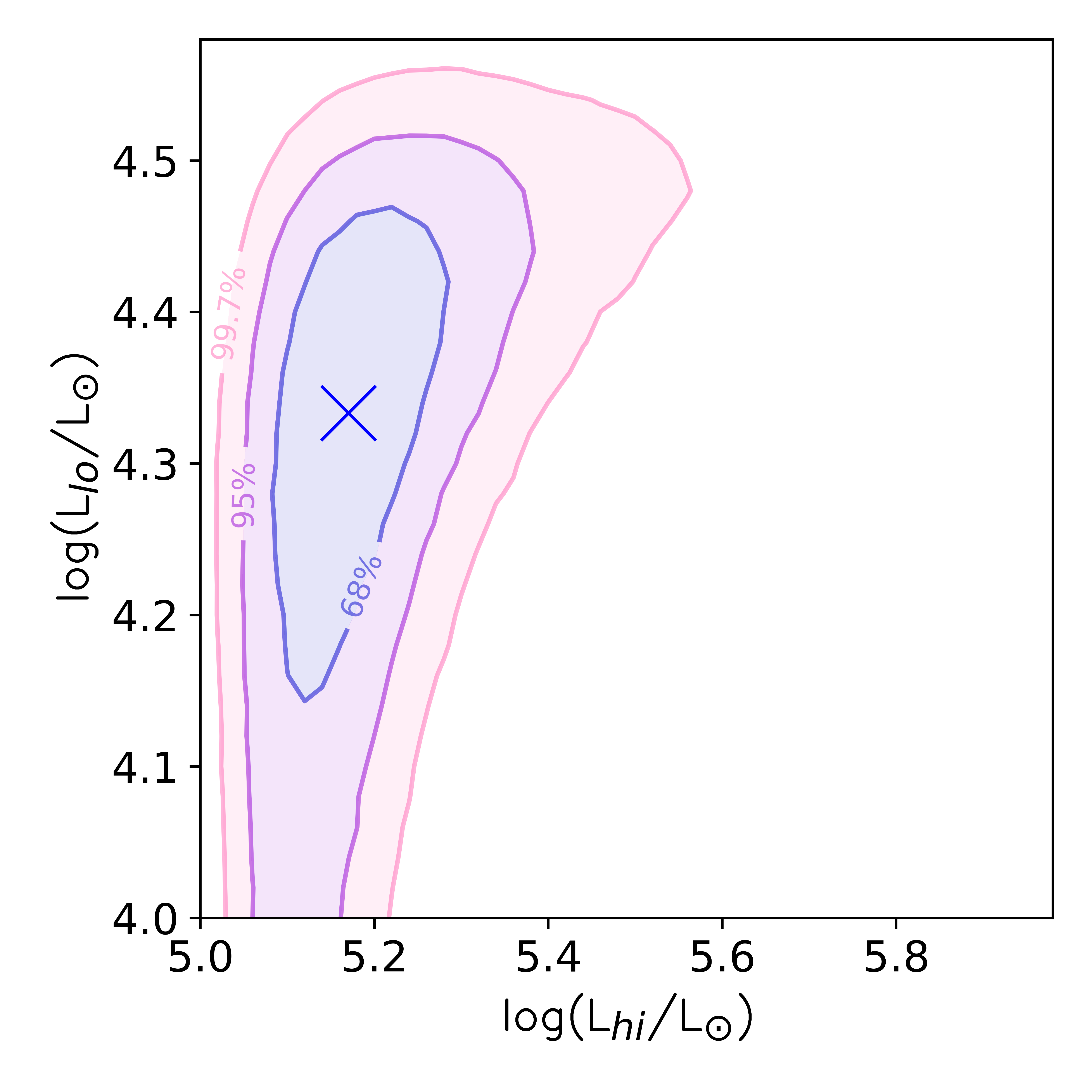}\includegraphics[width=\columnwidth]{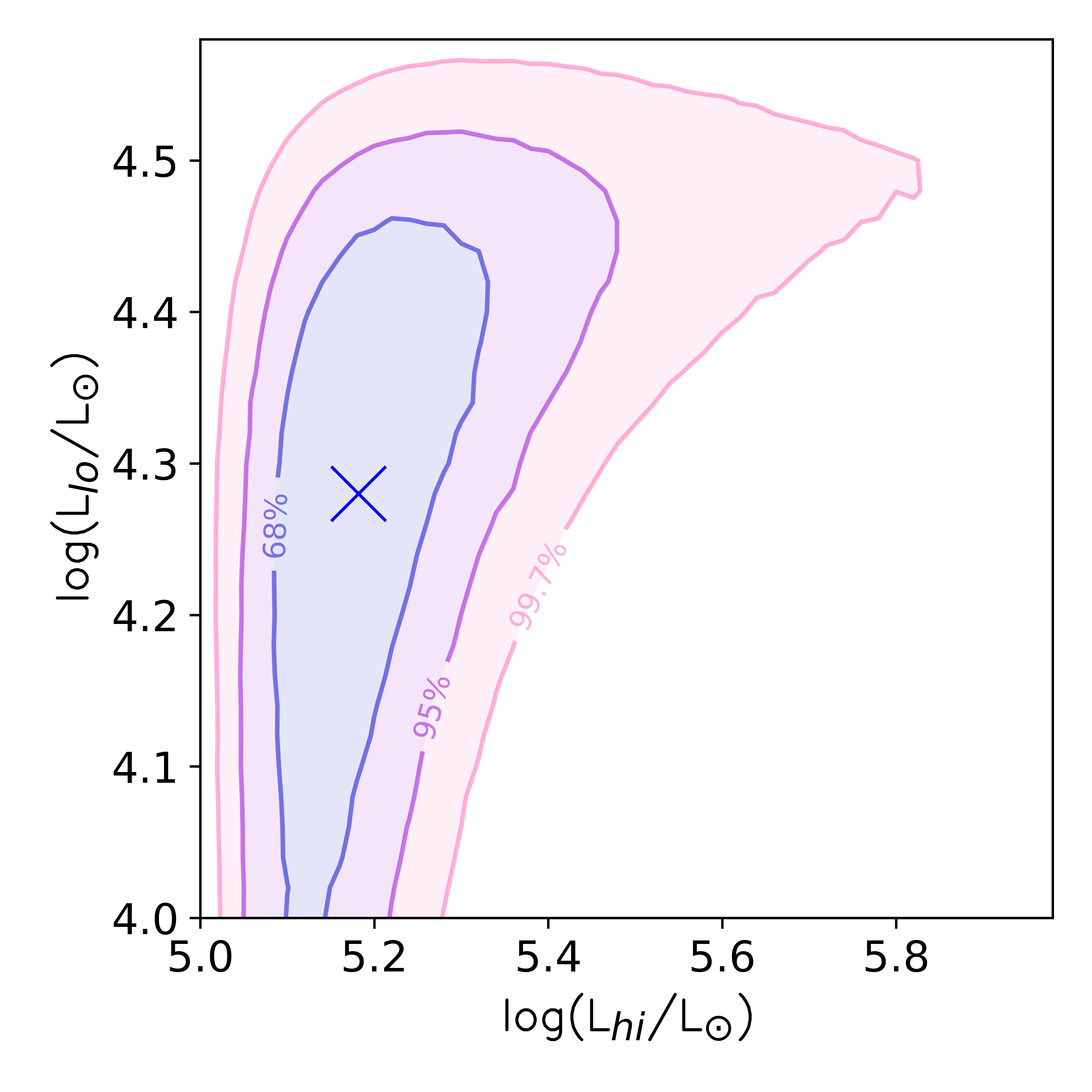}
    \caption{2D probability planes from the CLD fitting for the upper and lower ends of the luminosity distribution. The left-hand plot shows the results when considering the standard errors, while the right-hand plot shows the results when including an extra error on the F814W detection's of 0.2 dex. }
    \label{fig:lbols_with_err}
\end{figure*}

When including this additional uncertainty, the significance of the RSG problem is even more dubious. As demonstrated by \cite{davies2018initial} and \cite{davies2020on}, using later spectral types for progenitors yields higher luminosities/initial masses. Compared to the RSGs in M31, we show that when using the assumed BC from \cite{smartt2009death} (BCI=0), the luminosites of the RSGs in M31 are underestimated by an average of 0.3 dex. We now suggest that even if a later BC is assumed, the error bars on the derived luminosity are likely far larger than typically appreciated, due to unseen variability and an uncertain BC vs spectral type relation. In Fig \ref{fig:lbols_with_err} we show the constraints on the minimum $L_{lo}$ and maximum $L_{hi}$ progenitor luminosities within the 68, 95 and 99.7\% confidence limits. In the left-hand figure, we show the results when considering only the standard errors, while on the right we show the effect on the probability planes when including the larger errors. These figures show that it is not possible to rule out even the highest luminosity RSGs known (\logl\ $>$ 5.8) at the 3$\sigma$ level, see Fig. \ref{fig:lbols_with_err}. 

There is considerable additional model-dependent uncertainty when converting these luminosities to initial masses, so we have performed our analysis in terms of stellar luminosity.  However, for comparison with previous results, we now convert the upper luminosity limits to progenitor mass estimates using the same M-L relation as in \citet{smartt2009death}. For standard errors, we find $M_{hi}$ = 20, 23, 25 \msun \, at the 68, 95 and 99.7\% confidence levels respectively. When considering the appropriate extra error on the F814W luminosity estimates, $M_{hi}$ increases to 21, 24, and $>$ 25\msun\footnote{The M-L relation used previously does not extend as high as luminosities of \logl = 5.8.}. This upper mass limit is consistent with the observed upper mass limit for RSGs in local galaxies \citep[e.g.][]{davies2018humphreys,mcdonald2022red} and does not support the conclusion that high-mass progenitors are missing from the progenitor sample. Using Fig. 3 in \citet{smartt2015observational} we convert $L_{lo}$ to $M_{lo}$, finding a best fit value of 8\msun, slightly higher than the $M_{lo}$ found in previous studies. This result is consistent with the canonical theoretical predictions of stellar evolution. 

Figure \ref{fig:clds} also demonstrates the discrepancy between luminosity estimates for the three well-studied progenitor cases discussed in Section \ref{sec:midIR}. For each of these progenitors, we show the luminosity that is estimated from the full SED (including mid-IR photometry) with the filled blue circles. We also plot the luminosity that would have been found in each case if {\it only} F814W photometry were available, using both the BC correction for both early (BC=0.9) and late (BC=0) type RSGs, via the red open diamonds. In each of these cases the luminosity that would have been estimated from I-band photometry only is vastly underestimated compared to the SED luminosity, even when we consider the CLD with additional errors. Finally, we also plot the luminosity estimates for progenitors in the \citet{smartt2009death} study (i.e. using BC = 0.9) via the green symbols.  These are systematically lower than the revised luminosities (blue).

\section{Summary}

We have examined the population of RSGs in M31 in order to inform the ways that luminosities and initial masses are inferred for more distant SN progenitors from pre-explosion photometry.  To do this, we compare RSG luminosities measured from the full optical/IR SED to the luminosity inferred from a single filter detection and an assumed bolometriuc correction (BC).   This comparison yields two important effects:  (1) Luminosities inferred from single-band (usually I-band) detections are systematically lower than the full SED luminosity by about 0.3 dex, and (2) there is a significant spread in luminosity, larger than is accounted for in previous estimates of the uncertainty for SN progenitors.  Together, these two effects act to eliminate the RSG problem.

This confirms the results presented in \citet{davies2018initial} - that the bolometric corrections used in \citet{smartt2009death} and \citep{smartt2015observational} tend to underestimate the luminosity of SN progenitors. For the RSGs in M31, using bolometric corrections for later spectral types does reduce the systematic offset compared to SED-derived luminosities, but there is still dispersion between the SED-derived luminosities and the I-band-derived luminosities. We suggest that this dispersion is likely due to a range of unknown spectral types, extinction, and variability. When this additional error is taken into account for the sample of SN progenitors, the statistical significance of the RSG problem is further reduced.

We also use the three very nearby and well-studied examples of SN 2017eaw, SN 2023ixf and SN 2024ggi to demonstrate the systematic underestimation of luminosity that occurs when only I-band photometry is available, compared to an SED with mid-IR data points. In each of these cases, the luminosity derived from the more complete SED yielded a higher luminosity than when using only the I-band observation. For SN 2023ixf, this effect corresponded to a difference in luminosity of over 1.4 dex, or a difference in the implied initial mass of more than 10\msun. If we only had access to the I-band photometry, the luminosity would have been vastly underestimated. 

We therefore conclude that any progenitor luminosity derived without mid-IR data is likely to be highly uncertain, perhaps underestimated by as much as 1 dex, and should not be used in an attempt to constrain the progenitor luminosity function. Given the above results, we reiterate that the statistical evidence for the missing high-mass progenitors, the RSG problem, does not exist. 

Notably, these results also alleviate the tension at the lower end of the progenitor luminosity function, whereby some progenitors have been detected at luminosities that imply initial masses in the range of roughly 6-8 M$_{\odot}$.  This would have been problematic, since single stars in this initial mass range are too low mass to end their lives as CCSNe. This tension is eliminated if previously derived initial masses were systematically underestimated, as shown above.

\begin{acknowledgments}
The authors would like to thank the anonymous referee for constructive comments which helped improve the paper. We thank discussions with the community at the MIAPbP Interacting Supernovae workshop that motivated this work. This research was supported by the Munich Institute for Astro-, Particle and BioPhysics (MIAPbP) which is funded by the Deutsche Forschungsgemeinschaft (DFG, German Research Foundation) under Germany´s Excellence Strategy – EXC-2094 – 390783311. This work makes use of the IDL astrolib, as well as Python packages {\tt Astropy} \citep{astropy:2013, astropy:2018, astropy:2022}`, {\tt numpy} \citep{numpy}, {\tt pandas} \citep{pandas}, {\tt matplotlib} \citep{matplotlib}. 
\end{acknowledgments}

%





\bibliography{sample63}{}

\begin{thebibliography}{}
\expandafter\ifx\csname natexlab\endcsname\relax\def\natexlab#1{#1}\fi
\providecommand{\url}[1]{\href{#1}{#1}}
\providecommand{\dodoi}[1]{doi:~\href{http://doi.org/#1}{\nolinkurl{#1}}}
\providecommand{\doeprint}[1]{\href{http://ascl.net/#1}{\nolinkurl{http://ascl.net/#1}}}
\providecommand{\doarXiv}[1]{\href{https://arxiv.org/abs/#1}{\nolinkurl{https://arxiv.org/abs/#1}}}

\bibitem[{{Astropy Collaboration} {et~al.}(2013){Astropy Collaboration},
  {Robitaille}, {Tollerud}, {Greenfield}, {Droettboom}, {Bray}, {Aldcroft},
  {Davis}, {Ginsburg}, {Price-Whelan}, {Kerzendorf}, {Conley}, {Crighton},
  {Barbary}, {Muna}, {Ferguson}, {Grollier}, {Parikh}, {Nair}, {Unther},
  {Deil}, {Woillez}, {Conseil}, {Kramer}, {Turner}, {Singer}, {Fox}, {Weaver},
  {Zabalza}, {Edwards}, {Azalee Bostroem}, {Burke}, {Casey}, {Crawford},
  {Dencheva}, {Ely}, {Jenness}, {Labrie}, {Lim}, {Pierfederici}, {Pontzen},
  {Ptak}, {Refsdal}, {Servillat}, \& {Streicher}}]{astropy:2013}
{Astropy Collaboration}, {Robitaille}, T.~P., {Tollerud}, E.~J., {et~al.} 2013,
  \aap, 558, A33, \dodoi{10.1051/0004-6361/201322068}

\bibitem[{{Astropy Collaboration} {et~al.}(2018){Astropy Collaboration},
  {Price-Whelan}, {Sip{\H{o}}cz}, {G{\"u}nther}, {Lim}, {Crawford}, {Conseil},
  {Shupe}, {Craig}, {Dencheva}, {Ginsburg}, {Vand erPlas}, {Bradley},
  {P{\'e}rez-Su{\'a}rez}, {de Val-Borro}, {Aldcroft}, {Cruz}, {Robitaille},
  {Tollerud}, {Ardelean}, {Babej}, {Bach}, {Bachetti}, {Bakanov}, {Bamford},
  {Barentsen}, {Barmby}, {Baumbach}, {Berry}, {Biscani}, {Boquien}, {Bostroem},
  {Bouma}, {Brammer}, {Bray}, {Breytenbach}, {Buddelmeijer}, {Burke},
  {Calderone}, {Cano Rodr{\'\i}guez}, {Cara}, {Cardoso}, {Cheedella}, {Copin},
  {Corrales}, {Crichton}, {D'Avella}, {Deil}, {Depagne}, {Dietrich}, {Donath},
  {Droettboom}, {Earl}, {Erben}, {Fabbro}, {Ferreira}, {Finethy}, {Fox},
  {Garrison}, {Gibbons}, {Goldstein}, {Gommers}, {Greco}, {Greenfield},
  {Groener}, {Grollier}, {Hagen}, {Hirst}, {Homeier}, {Horton}, {Hosseinzadeh},
  {Hu}, {Hunkeler}, {Ivezi{\'c}}, {Jain}, {Jenness}, {Kanarek}, {Kendrew},
  {Kern}, {Kerzendorf}, {Khvalko}, {King}, {Kirkby}, {Kulkarni}, {Kumar},
  {Lee}, {Lenz}, {Littlefair}, {Ma}, {Macleod}, {Mastropietro}, {McCully},
  {Montagnac}, {Morris}, {Mueller}, {Mumford}, {Muna}, {Murphy}, {Nelson},
  {Nguyen}, {Ninan}, {N{\"o}the}, {Ogaz}, {Oh}, {Parejko}, {Parley}, {Pascual},
  {Patil}, {Patil}, {Plunkett}, {Prochaska}, {Rastogi}, {Reddy Janga},
  {Sabater}, {Sakurikar}, {Seifert}, {Sherbert}, {Sherwood-Taylor}, {Shih},
  {Sick}, {Silbiger}, {Singanamalla}, {Singer}, {Sladen}, {Sooley},
  {Sornarajah}, {Streicher}, {Teuben}, {Thomas}, {Tremblay}, {Turner},
  {Terr{\'o}n}, {van Kerkwijk}, {de la Vega}, {Watkins}, {Weaver}, {Whitmore},
  {Woillez}, {Zabalza}, \& {Astropy Contributors}}]{astropy:2018}
{Astropy Collaboration}, {Price-Whelan}, A.~M., {Sip{\H{o}}cz}, B.~M., {et~al.}
  2018, \aj, 156, 123, \dodoi{10.3847/1538-3881/aabc4f}

\bibitem[{{Astropy Collaboration} {et~al.}(2022){Astropy Collaboration},
  {Price-Whelan}, {Lim}, {Earl}, {Starkman}, {Bradley}, {Shupe}, {Patil},
  {Corrales}, {Brasseur}, {N{"o}the}, {Donath}, {Tollerud}, {Morris},
  {Ginsburg}, {Vaher}, {Weaver}, {Tocknell}, {Jamieson}, {van Kerkwijk},
  {Robitaille}, {Merry}, {Bachetti}, {G{"u}nther}, {Aldcroft},
  {Alvarado-Montes}, {Archibald}, {B{'o}di}, {Bapat}, {Barentsen}, {Baz{'a}n},
  {Biswas}, {Boquien}, {Burke}, {Cara}, {Cara}, {Conroy}, {Conseil}, {Craig},
  {Cross}, {Cruz}, {D'Eugenio}, {Dencheva}, {Devillepoix}, {Dietrich},
  {Eigenbrot}, {Erben}, {Ferreira}, {Foreman-Mackey}, {Fox}, {Freij}, {Garg},
  {Geda}, {Glattly}, {Gondhalekar}, {Gordon}, {Grant}, {Greenfield}, {Groener},
  {Guest}, {Gurovich}, {Handberg}, {Hart}, {Hatfield-Dodds}, {Homeier},
  {Hosseinzadeh}, {Jenness}, {Jones}, {Joseph}, {Kalmbach}, {Karamehmetoglu},
  {Ka{l}uszy{'n}ski}, {Kelley}, {Kern}, {Kerzendorf}, {Koch}, {Kulumani},
  {Lee}, {Ly}, {Ma}, {MacBride}, {Maljaars}, {Muna}, {Murphy}, {Norman},
  {O'Steen}, {Oman}, {Pacifici}, {Pascual}, {Pascual-Granado}, {Patil},
  {Perren}, {Pickering}, {Rastogi}, {Roulston}, {Ryan}, {Rykoff}, {Sabater},
  {Sakurikar}, {Salgado}, {Sanghi}, {Saunders}, {Savchenko}, {Schwardt},
  {Seifert-Eckert}, {Shih}, {Jain}, {Shukla}, {Sick}, {Simpson},
  {Singanamalla}, {Singer}, {Singhal}, {Sinha}, {Sip{H{o}}cz}, {Spitler},
  {Stansby}, {Streicher}, {{{S}}umak}, {Swinbank}, {Taranu}, {Tewary},
  {Tremblay}, {Val-Borro}, {Van Kooten}, {Vasovi{'c}}, {Verma}, {de Miranda
  Cardoso}, {Williams}, {Wilson}, {Winkel}, {Wood-Vasey}, {Xue}, {Yoachim},
  {Zhang}, {Zonca}, \& {Astropy Project Contributors}}]{astropy:2022}
{Astropy Collaboration}, {Price-Whelan}, A.~M., {Lim}, P.~L., {et~al.} 2022,
  \apj, 935, 167, \dodoi{10.3847/1538-4357/ac7c74}

\bibitem[{{Beasor} \& {Smith}(2022)}]{beasor2022dusty}
{Beasor}, E.~R., \& {Smith}, N. 2022, \apj, 933, 41,
  \dodoi{10.3847/1538-4357/ac6dcf}

\bibitem[{{Dalcanton} {et~al.}(2012){Dalcanton}, {Williams}, {Lang}, {Lauer},
  {Kalirai}, {Seth}, {Dolphin}, {Rosenfield}, {Weisz}, {Bell}, {Bianchi},
  {Boyer}, {Caldwell}, {Dong}, {Dorman}, {Gilbert}, {Girardi}, {Gogarten},
  {Gordon}, {Guhathakurta}, {Hodge}, {Holtzman}, {Johnson}, {Larsen}, {Lewis},
  {Melbourne}, {Olsen}, {Rix}, {Rosema}, {Saha}, {Sarajedini}, {Skillman}, \&
  {Stanek}}]{dalcanton2012ophat}
{Dalcanton}, J.~J., {Williams}, B.~F., {Lang}, D., {et~al.} 2012, \apjs, 200,
  18, \dodoi{10.1088/0067-0049/200/2/18}

\bibitem[{Davies \& Beasor(2018)}]{davies2018initial}
Davies, B., \& Beasor, E.~R. 2018, Monthly Notices of the Royal Astronomical
  Society, 474, 2116

\bibitem[{{Davies} \& {Beasor}(2020{\natexlab{a}})}]{davies2020red}
{Davies}, B., \& {Beasor}, E.~R. 2020{\natexlab{a}}, \mnras, 493, 468,
  \dodoi{10.1093/mnras/staa174}

\bibitem[{{Davies} \& {Beasor}(2020{\natexlab{b}})}]{davies2020on}
---. 2020{\natexlab{b}}, \mnras, 496, L142, \dodoi{10.1093/mnrasl/slaa102}

\bibitem[{Davies {et~al.}(2018)Davies, Crowther, \&
  Beasor}]{davies2018humphreys}
Davies, B., Crowther, P.~A., \& Beasor, E.~R. 2018, Monthly Notices of the
  Royal Astronomical Society, 478, 3138

\bibitem[{Elias {et~al.}(1985)Elias, Frogel, \& Humphreys}]{elias1985m}
Elias, J., Frogel, J., \& Humphreys, R. 1985, The Astrophysical Journal
  Supplement Series, 57, 91

\bibitem[{{Fraser} {et~al.}(2014){Fraser}, {Maund}, {Smartt}, {Kotak},
  {Lawrence}, {Bruce}, {Valenti}, {Yuan}, {Benetti}, {Chen}, {Gal-Yam},
  {Inserra}, \& {Young}}]{fraser2014ej}
{Fraser}, M., {Maund}, J.~R., {Smartt}, S.~J., {et~al.} 2014, \mnras, 439, L56,
  \dodoi{10.1093/mnrasl/slt179}

\bibitem[{Georgy(2012)}]{georgy2012yellow}
Georgy, C. 2012, Astronomy \& Astrophysics, 538, L8

\bibitem[{Harris {et~al.}(2020)Harris, Millman, van~der Walt, Gommers,
  Virtanen, Cournapeau, Wieser, Taylor, Berg, Smith, Kern, Picus, Hoyer, van
  Kerkwijk, Brett, Haldane, del R{\'{i}}o, Wiebe, Peterson,
  G{\'{e}}rard-Marchant, Sheppard, Reddy, Weckesser, Abbasi, Gohlke, \&
  Oliphant}]{numpy}
Harris, C.~R., Millman, K.~J., van~der Walt, S.~J., {et~al.} 2020, Nature, 585,
  357, \dodoi{10.1038/s41586-020-2649-2}

\bibitem[{{Heger} {et~al.}(2003){Heger}, {Fryer}, {Woosley}, {Langer}, \&
  {Hartmann}}]{heger2003how}
{Heger}, A., {Fryer}, C.~L., {Woosley}, S.~E., {Langer}, N., \& {Hartmann},
  D.~H. 2003, \apj, 591, 288, \dodoi{10.1086/375341}

\bibitem[{{Humphreys} {et~al.}(1988){Humphreys}, {Pennington}, {Jones}, \&
  {Ghigo}}]{humphreys1988m31}
{Humphreys}, R.~M., {Pennington}, R.~L., {Jones}, T.~J., \& {Ghigo}, F.~D.
  1988, \aj, 96, 1884, \dodoi{10.1086/114935}

\bibitem[{Hunter(2007)}]{matplotlib}
Hunter, J.~D. 2007, Computing in Science \& Engineering, 9, 90,
  \dodoi{10.1109/MCSE.2007.55}

\bibitem[{Itagaki(2023)}]{itagaki2023}
Itagaki, K. 2023, Transient Name Server Discovery Report, 2023-1158, 1.
\newblock \url{https://ui.adsabs.harvard.edu/abs/2023TNSTR1158....1I}

\bibitem[{{Jacobson-Gal{\'a}n} {et~al.}(2022){Jacobson-Gal{\'a}n}, {Dessart},
  {Jones}, {Margutti}, {Coppejans}, {Dimitriadis}, {Foley}, {Kilpatrick},
  {Matthews}, {Rest}, {Terreran}, {Aleo}, {Auchettl}, {Blanchard}, {Coulter},
  {Davis}, {de Boer}, {DeMarchi}, {Drout}, {Earl}, {Gagliano}, {Gall},
  {Hjorth}, {Huber}, {Ibik}, {Milisavljevic}, {Pan}, {Rest}, {Ridden-Harper},
  {Rojas-Bravo}, {Siebert}, {Smith}, {Taggart}, {Tinyanont}, {Wang}, \&
  {Zenati}}]{jacobson2022final}
{Jacobson-Gal{\'a}n}, W.~V., {Dessart}, L., {Jones}, D.~O., {et~al.} 2022,
  \apj, 924, 15, \dodoi{10.3847/1538-4357/ac3f3a}

\bibitem[{{Jencson} {et~al.}(2023){Jencson}, {Pearson}, {Beasor}, {Lau},
  {Andrews}, {Bostroem}, {Dong}, {Engesser}, {Gomez}, {Guolo}, {Hoang},
  {Hosseinzadeh}, {Jha}, {Karambelkar}, {Kasliwal}, {Lundquist}, {Meza
  Retamal}, {Rest}, {Sand}, {Shahbandeh}, {Shrestha}, {Smith}, {Strader},
  {Valenti}, {Wang}, \& {Zenati}}]{jencson2023}
{Jencson}, J.~E., {Pearson}, J., {Beasor}, E.~R., {et~al.} 2023, \apjl, 952,
  L30, \dodoi{10.3847/2041-8213/ace618}

\bibitem[{{Karachentsev} {et~al.}(2004){Karachentsev}, {Karachentseva},
  {Huchtmeier}, \& {Makarov}}]{karachentsev2004m31}
{Karachentsev}, I.~D., {Karachentseva}, V.~E., {Huchtmeier}, W.~K., \&
  {Makarov}, D.~I. 2004, \aj, 127, 2031, \dodoi{10.1086/382905}

\bibitem[{{Kilpatrick} \& {Foley}(2018)}]{kilpatrick2018}
{Kilpatrick}, C.~D., \& {Foley}, R.~J. 2018, \mnras, 481, 2536,
  \dodoi{10.1093/mnras/sty2435}

\bibitem[{{Kilpatrick} {et~al.}(2023){Kilpatrick}, {Foley},
  {Jacobson-Gal{\'a}n}, {Piro}, {Smartt}, {Drout}, {Gagliano}, {Gall},
  {Hjorth}, {Jones}, {Mandel}, {Margutti}, {Ramirez-Ruiz}, {Ransome}, {Villar},
  {Coulter}, {Gao}, {Matthews}, {Taggart}, \& {Zenati}}]{kilpatrick2023}
{Kilpatrick}, C.~D., {Foley}, R.~J., {Jacobson-Gal{\'a}n}, W.~V., {et~al.}
  2023, \apjl, 952, L23, \dodoi{10.3847/2041-8213/ace4ca}

\bibitem[{{Kochanek}(2020)}]{kochanek2020on}
{Kochanek}, C.~S. 2020, \mnras, 493, 4945, \dodoi{10.1093/mnras/staa605}

\bibitem[{{Massey} {et~al.}(2016){Massey}, {Neugent}, \&
  {Smart}}]{massey2016m31}
{Massey}, P., {Neugent}, K.~F., \& {Smart}, B.~M. 2016, \aj, 152, 62,
  \dodoi{10.3847/0004-6256/152/3/62}

\bibitem[{Maund {et~al.}(2004)Maund, Smartt, Kudritzki, Podsiadlowski, \&
  Gilmore}]{maund2004massive}
Maund, J.~R., Smartt, S.~J., Kudritzki, R.~P., Podsiadlowski, P., \& Gilmore,
  G.~F. 2004, Nature, 427, 129

\bibitem[{{McDonald} {et~al.}(2022){McDonald}, {Davies}, \&
  {Beasor}}]{mcdonald2022red}
{McDonald}, S. L.~E., {Davies}, B., \& {Beasor}, E.~R. 2022, \mnras, 510, 3132,
  \dodoi{10.1093/mnras/stab3453}

\bibitem[{{Meynet} \& {Maeder}(2000)}]{meynet2000stellar}
{Meynet}, G., \& {Maeder}, A. 2000, \aap, 361, 101

\bibitem[{{Neustadt} {et~al.}(2024){Neustadt}, {Kochanek}, \&
  {Smith}}]{neustadt2023}
{Neustadt}, J.~M.~M., {Kochanek}, C.~S., \& {Smith}, M.~R. 2024, \mnras, 527,
  5366, \dodoi{10.1093/mnras/stad3073}

\bibitem[{{Niu} {et~al.}(2023){Niu}, {Sun}, {Maund}, {Zhang}, {Zhao}, \&
  {Liu}}]{niu2023}
{Niu}, Z., {Sun}, N.-C., {Maund}, J.~R., {et~al.} 2023, \apjl, 955, L15,
  \dodoi{10.3847/2041-8213/acf4e3}

\bibitem[{{Pledger} \& {Shara}(2023)}]{pledger2023}
{Pledger}, J.~L., \& {Shara}, M.~M. 2023, \apjl, 953, L14,
  \dodoi{10.3847/2041-8213/ace88b}

\bibitem[{{Qin} {et~al.}(2023){Qin}, {Zhang}, {Bloom}, {Sollerman},
  {Zimmerman}, {Irani}, {Schulze}, {Gal-Yam}, {Kasliwal}, {Coughlin}, {Perley},
  {Fremling}, \& {Kulkarni}}]{qin2023}
{Qin}, Y.-J., {Zhang}, K., {Bloom}, J., {et~al.} 2023, arXiv e-prints,
  arXiv:2309.10022, \dodoi{10.48550/arXiv.2309.10022}

\bibitem[{{Ransome} {et~al.}(2024){Ransome}, {Villar}, {Tartaglia}, {Gonzalez},
  {Jacobson-Gal{\'a}n}, {Kilpatrick}, {Margutti}, {Foley}, {Grayling}, {Ni},
  {Yarza}, {Ye}, {Auchettl}, {de Boer}, {Chambers}, {Coulter}, {Drout},
  {Farias}, {Gall}, {Gao}, {Huber}, {Ibik}, {Jones}, {Khetan}, {Lin},
  {Politsch}, {Raimundo}, {Rest}, {Wainscoat}, {Yadavalli}, \&
  {Zenati}}]{ransome2024}
{Ransome}, C.~L., {Villar}, V.~A., {Tartaglia}, A., {et~al.} 2024, \apj, 965,
  93, \dodoi{10.3847/1538-4357/ad2df7}

\bibitem[{{Riess} {et~al.}(2022){Riess}, {Yuan}, {Macri}, {Scolnic}, {Brout},
  {Casertano}, {Jones}, {Murakami}, {Anand}, {Breuval}, {Brink}, {Filippenko},
  {Hoffmann}, {Jha}, {D'arcy Kenworthy}, {Mackenty}, {Stahl}, \&
  {Zheng}}]{riess2022}
{Riess}, A.~G., {Yuan}, W., {Macri}, L.~M., {et~al.} 2022, \apjl, 934, L7,
  \dodoi{10.3847/2041-8213/ac5c5b}

\bibitem[{Smartt(2015)}]{smartt2015observational}
Smartt, S. 2015, Publications of the Astronomical Society of Australia, 32,
  e016

\bibitem[{Smartt {et~al.}(2009)Smartt, Eldridge, Crockett, \&
  Maund}]{smartt2009death}
Smartt, S., Eldridge, J., Crockett, R., \& Maund, J.~R. 2009, Monthly Notices
  of the Royal Astronomical Society, 395, 1409

\bibitem[{{Smith} {et~al.}(2009){Smith}, {Hinkle}, \& {Ryde}}]{smith09vy}
{Smith}, N., {Hinkle}, K.~H., \& {Ryde}, N. 2009, \aj, 137, 3558,
  \dodoi{10.1088/0004-6256/137/3/3558}

\bibitem[{Smith {et~al.}(2011)Smith, Li, Filippenko, \&
  Chornock}]{smith2011observed}
Smith, N., Li, W., Filippenko, A.~V., \& Chornock, R. 2011, Monthly Notices of
  the Royal Astronomical Society, 412, 1522

\bibitem[{{Soraisam} {et~al.}(2018){Soraisam}, {Bildsten}, {Drout}, {Bauer},
  {Gilfanov}, {Kupfer}, {Laher}, {Masci}, {Prince}, {Kulkarni}, {Matheson}, \&
  {Saha}}]{soraisam2018}
{Soraisam}, M.~D., {Bildsten}, L., {Drout}, M.~R., {et~al.} 2018, \apj, 859,
  73, \dodoi{10.3847/1538-4357/aabc59}

\bibitem[{{Soraisam} {et~al.}(2023){Soraisam}, {Szalai}, {Van Dyk}, {Andrews},
  {Srinivasan}, {Chun}, {Matheson}, {Scicluna}, \&
  {Vasquez-Torres}}]{soraisam2023}
{Soraisam}, M.~D., {Szalai}, T., {Van Dyk}, S.~D., {et~al.} 2023, \apj, 957,
  64, \dodoi{10.3847/1538-4357/acef22}

\bibitem[{{Sukhbold} {et~al.}(2018){Sukhbold}, {Woosley}, \&
  {Heger}}]{sukhbold2018high}
{Sukhbold}, T., {Woosley}, S.~E., \& {Heger}, A. 2018, \apj, 860, 93,
  \dodoi{10.3847/1538-4357/aac2da}

\bibitem[{{Tikhonov}(2014)}]{tikhonov2014}
{Tikhonov}, N.~A. 2014, Astronomy Letters, 40, 537,
  \dodoi{10.1134/S1063773714090035}

\bibitem[{{Tonry} {et~al.}(2024){Tonry}, {Denneau}, {Weiland}, {Lawrence},
  {Siverd}, {Erasmus}, {Koorts}, {Jordan}, {Suc}, {Smartt}, {Smith}, {Young},
  {Nicholl}, {Fulton}, {McCollum}, {Moore}, {Weston}, {Sheng}, {Ramsden},
  {Angus}, {Aamer}, {Shingles}, {Srivastav}, {Gillanders}, {Rhodes},
  {Andersson}, {Stevance}, {Rest}, {Chen}, {Stubbs}, \& {Sommer}}]{tonry2024}
{Tonry}, J., {Denneau}, L., {Weiland}, H., {et~al.} 2024, Transient Name Server
  Discovery Report, 2024-1020, 1

\bibitem[{{Tully} {et~al.}(2013){Tully}, {Courtois}, {Dolphin}, {Fisher},
  {H{\'e}raudeau}, {Jacobs}, {Karachentsev}, {Makarov}, {Makarova},
  {Mitronova}, {Rizzi}, {Shaya}, {Sorce}, \& {Wu}}]{tully2013}
{Tully}, R.~B., {Courtois}, H.~M., {Dolphin}, A.~E., {et~al.} 2013, \aj, 146,
  86, \dodoi{10.1088/0004-6256/146/4/86}

\bibitem[{{Van Dyk} {et~al.}(2019){Van Dyk}, {Zheng}, {Maund}, {Brink},
  {Srinivasan}, {Andrews}, {Smith}, {Leonard}, {Morozova}, {Filippenko},
  {Conner}, {Milisavljevic}, {de Jaeger}, {Long}, {Isaacson}, {Crossfield},
  {Kosiarek}, {Howard}, {Fox}, {Kelly}, {Piro}, {Littlefair}, {Dhillon},
  {Wilson}, {Butterley}, {Yunus}, {Channa}, {Jeffers}, {Falcon}, {Ross},
  {Hestenes}, {Stegman}, {Zhang}, \& {Kumar}}]{vandyk2019}
{Van Dyk}, S.~D., {Zheng}, W., {Maund}, J.~R., {et~al.} 2019, \apj, 875, 136,
  \dodoi{10.3847/1538-4357/ab1136}

\bibitem[{{Van Dyk} {et~al.}(2024){Van Dyk}, {Srinivasan}, {Andrews},
  {Soraisam}, {Szalai}, {Howell}, {Isaacson}, {Matheson}, {Petigura},
  {Scicluna}, {Stephens}, {Van Zandt}, {Zheng}, {Chun}, \&
  {Fillippenko}}]{vandyk2024}
{Van Dyk}, S.~D., {Srinivasan}, S., {Andrews}, J.~E., {et~al.} 2024, \apj, 968,
  27, \dodoi{10.3847/1538-4357/ad414b}

\bibitem[{Walmswell \& Eldridge(2012)}]{walmswell2012circumstellar}
Walmswell, J.~J., \& Eldridge, J.~J. 2012, Monthly Notices of the Royal
  Astronomical Society, 419, 2054

\bibitem[{{W}es {M}c{K}inney(2010)}]{pandas}
{W}es {M}c{K}inney. 2010, in {P}roceedings of the 9th {P}ython in {S}cience
  {C}onference, ed. {S}t\'efan van~der {W}alt \& {J}arrod {M}illman, 56 -- 61,
  \dodoi{10.25080/Majora-92bf1922-00a}

\bibitem[{{Xiang} {et~al.}(2024{\natexlab{a}}){Xiang}, {Mo}, {Wang}, {Wang},
  {Zhang}, {Lin}, \& {Wang}}]{xiang2024a}
{Xiang}, D., {Mo}, J., {Wang}, L., {et~al.} 2024{\natexlab{a}}, Science China
  Physics, Mechanics, and Astronomy, 67, 219514,
  \dodoi{10.1007/s11433-023-2267-0}

\bibitem[{{Xiang} {et~al.}(2024{\natexlab{b}}){Xiang}, {Mo}, {Wang}, {Wang},
  {Zhang}, {Lin}, {Chen}, {Song}, {Liu}, {Wang}, \& {Li}}]{xiang2024b}
{Xiang}, D., {Mo}, J., {Wang}, X., {et~al.} 2024{\natexlab{b}}, arXiv e-prints,
  arXiv:2405.07699, \dodoi{10.48550/arXiv.2405.07699}

\bibitem[{{Zapartas} {et~al.}(2021){Zapartas}, {de Mink}, {Justham}, {Smith},
  {Renzo}, \& {de Koter}}]{zapartas2021effect}
{Zapartas}, E., {de Mink}, S.~E., {Justham}, S., {et~al.} 2021, \aap, 645, A6,
  \dodoi{10.1051/0004-6361/202037744}

\bibitem[{{Zapartas} {et~al.}(2017){Zapartas}, {de Mink}, {Izzard}, {Yoon},
  {Badenes}, {G{\"o}tberg}, {de Koter}, {Neijssel}, {Renzo}, {Schootemeijer},
  \& {Shrotriya}}]{zapartas2017}
{Zapartas}, E., {de Mink}, S.~E., {Izzard}, R.~G., {et~al.} 2017, \aap, 601,
  A29, \dodoi{10.1051/0004-6361/201629685}

\bibitem[{{Zapartas} {et~al.}(2019){Zapartas}, {de Mink}, {Justham}, {Smith},
  {de Koter}, {Renzo}, {Arcavi}, {Farmer}, {G{\"o}tberg}, \&
  {Toonen}}]{zapartas2019}
{Zapartas}, E., {de Mink}, S.~E., {Justham}, S., {et~al.} 2019, \aap, 631, A5,
  \dodoi{10.1051/0004-6361/201935854}

\bibitem[{{Zhang} {et~al.}(2024){Zhang}, {Dessart}, {Wang}, {Zhai}, {Yang},
  {Li}, {Lin}, {Valerin}, {Cai}, {Guo}, {Wang}, {Zhao}, {Wang}, \&
  {Yan}}]{zhang2024}
{Zhang}, J., {Dessart}, L., {Wang}, X., {et~al.} 2024, arXiv e-prints,
  arXiv:2406.07806, \dodoi{10.48550/arXiv.2406.07806}

\end{thebibliography}
\bibliographystyle{aasjournal}



\end{document}